# Design, Implementation, and Performance Evaluation of a Fiber Bragg Gratings (FBG) based Smart Insole to Measure Plantar Pressure and Temperature

Sakib Mahmud[1], Amith Khandakar[1,2], Muhammad E. H. Chowdhury[1*], Mohammed AbdulMoniem[1], Mamun Bin Ibne Reaz[2], Zaid Bin Mahbub[3], Kishor Kumar Sadasivuni[4], M Murugappan[5,6], Mohammed Alhatou[7]

[1]Department of Electrical Engineering, Qatar University, Doha-2713, Qatar
[2]Dept. of Electrical, Electronics and Systems Engineering, Universiti Kebangsaan Malaysia, Bangi, Selangor 43600, Malaysia
[3]Department of Physics and Mathematics, North South University, Dhaka, Bangladesh
[4]Center for Advanced Materials, Qatar University, Qatar
[5]Intelligent Signal Processing (ISP) Research Lab, Department of Electronics and Communication Engineering, Kuwait College of Science and Technology, Block 4, Doha, 13133, Kuwait
[6]Department of Electronics and Communication Engineering, School of Engineering, Vels Institute of Sciences, Technology, and Advanced Studies, Chennai, Tamil Nadu, India
[7]Neuromuscular Division, Hamad General Hospital and Department of Neurology; Alkhor Hospital, Doha, 3050, Qatar

***Correspondence:** Muhammad E. H. Chowdhury (mchowdhury@qu.edu.qa)

Abstract

Various foot complications can be easily avoided by continuous monitoring of foot plantar temperature and pressure. In this work, the design, characterization, and implementation of a Fiber Bragg Gratings (FBG) based smart insole capable of simultaneously measuring plantar pressure and temperature has been reported. The instrumented insole was tested and verified during static and gait exercises. The paper also provides a comparison of the developed optoelectronic-based solution with a commercially available and widely used plantar pressure measurement and analysis system and a lab-made, electronic sensor-based smart insole that can measure both plantar temperature and pressure. It was shown that even though the commercially available expensive system is very robust and highly precise due to many sensing units on the insole, the developed insole with a smaller number of sensors can provide both plantar temperature and pressure with reasonable precision while displaying both foot pressure and temperature maps, and gait cycle plots in real-time. The FBG-based solution is comfortable and safer than the other systems being compared. Thus, confirming the effectiveness of the proposed solution as an addition to the research area for detecting foot complications non-invasively using foot insoles.

## 1. Introduction

The advancement of electronics and sensors has helped in the development of a more portable and reliable diagnostic wearable for the early detection of anomalies [1]. The emphasis on e-Health, especially during the pandemic period, has driven the need for self-diagnosis solutions (using sensor systems with machine learning solutions) and helps the medical staff in the early diagnosis of anomalies without compromising their daily routines [2-4]. Self-diagnosis at home, i.e., self-care, which means monitoring without medical assistance, could be useful in preventing severe after-effects. Health Complications that can be early diagnosed from foot depend a lot on visual inspection which has its limitation for people suffering from obesity or visual impairment. It has also been confirmed that patients going through continuous monitoring of their foot temperature had a low risk of foot complications [5, 6]. As early as the 1970s, skin-temperature monitoring emerged as a useful tool for identifying patients at risk for ulceration. In literature, the temperature monitoring approach used plantar foot temperature asymmetry between the pair of feet, which is referred to as "asymmetry analysis", to find ulcers at an early stage [5-7]. Thus, we need a sensor-based solution that can correctly detect these temperature changes. Similar to temperature difference, the plantar pressure also provides information regarding diabetic foot complications. Nahas et al. in [8] have confirmed the strong correlation between plantar temperature and pressure distribution in a study involving 25 healthy patients. Later, Deschamps et al. in [9], with the help of statistical analysis and clustering of plantar pressure maps using k-mean clustering, confirmed the distinguishability between Healthy and Diabetic patients. Thus, improved sensors to provide these temperature and pressure maps can help in monitoring the progression of diabetic foot complications.

Several tools to measure plantar pressure and temperatures are available with various limitations on one or both quantities. For example, foot temperature measurement units are normally limited to a single measurement per day or designed for use only under clinical supervision [10]. Often the restrictions arise due to the specific purpose the system is developed e.g., early detection of Diabetic Foot Ulcer (DFU). The solutions proposed in [11, 12] use an infrared, handheld thermometer for measuring and comparing the temperature at six locations on both feet each morning. The design used the threshold temperature difference of 4°F (2.22°C) or higher [13, 14] as an early sign of DFU. Limitations can also be caused due to the inflexibility of the developed system in capturing data from a wide range of human sizes and shapes. For example, the systems developed in [11] can also result in false alarms, as it can be subjective to manually measure the temperature at different locations of the foot, especially when the feet have different sizes and shapes [13]. Another solution used a "smart mat" for measuring the foot temperature daily [14]. The nature of the temperature variations on both of the feet could be used to find the locations with higher temperatures and thus identify the formation of potential ulcers at an initial stage. There are some innovative wearables such as "smart socks", and "smart insole" with embedded sensors for measuring temperature [13, 15, 16]. Apart from its usage in acquiring gait cycle and relevant parameters, plantar pressure is also crucial for avoiding false temperature asymmetry alarms. Other than excessive cost, current commercially available solutions are also limited in terms of acquiring foot pressure and temperature together in real-time. Based on our literature review, there is no commercially available optoelectronic sensor-based smart insole. These drawbacks not only limit the current systems to convey the whole picture of the specific complexity it has been built to monitor for, but also restrict their generality while used for multiple applications. Optoelectronic-based Fiber Bragg Gratings (FBG) have recently garnered a lot of popularity in biomedical applications due to their immunity to electromagnetic interference, multiplexing opportunity, no extra requirement for humidity compensation, and high accuracy, and resolution [17, 18]. FBG is used as a temperature sensor due to its non-interfering nature to the radio frequency pattern. They have also been used as pressure transducers for intravascular measurements of pressure variations [19], radial artery pulse waveforms [20], respiratory plethysmography [21], hand-exoskeleton interactive force detection [22], for Gait plantar and shear force monitoring [18].

Some studies have been published during the past few years which focused on using optical-sensor-based solutions for developing smart wearables to identify and monitor various foot complications. There are socks made entirely of optical fiber [23], which showed great promise in accuracy but has a huge drawback due to the fragility of the optical fiber while wearing the sock [12]. This fragility issue can be solved by using hard insoles instead of socks, as explored in most studies. Leal-Junior et al. in their 2018 study [24] developed an intensity variation-based Polymer Optical Fiber (POF) system for acquiring knee joint angle responses from POF curvature sensors and gait cycle from a smart insole containing four POF sensors, which can be used for gait assistance and rehabilitation. Lakho et al. [25] developed an FBG-based insole containing four FBG sensors for vertical plantar pressure measurement and presented their system outcomes through different activities to prove the system's effectiveness in understanding body posture changes. One research group from Portugal performed several studies to understand the effectiveness of using FBG-based smart insoles in monitoring and treating patients with foot complications from different aspects. In one of their studies, Domingues et al. [26, 27] proposed an e-health solution consisting of a Fiber Bragg Gratings (FBG) based smart insole for plantar pressure data acquisition, and a wireless transceiver for transmitting the gait-cycle information along with data from a few other sensors to a remote server connected to the cloud for facilitating real-time patient monitoring and health management through mobile applications. The insole developed in [27] contained six FBG sensors inscribed into a cork-based structure for gait cycle measurement, along with a temperature sensor for foot temperature monitoring and compensation during data acquisition. In this study from the same group [28], they have shown the smart insole development in detail, from insole development to sensor placement, FBG sensor characterization, setting up data pipeline for a cloud-based solution, etc. The developed solution had four FBG sensors spread intelligently around the insole surface. In another of their study [29], they proposed a complete gait monitoring solution consisting of an insole for gait cycle measurements and two additional systems for concurrently monitoring knee and ankle movements using FBG. Each insole contained four FBG sensing units while knee and ankle movements were monitored using a single FBG sensor. In one of their more recent papers, Tavares et al. [30] analyzed both vertical and shear plantar force from gait-cycle data acquired by an FBG sensor-based insole where comparison was shown between optical and electronic sensor-based systems and a strong correlation could be observed for both vertical and shear forces. In this study, an FBG-based temperature sensor was used to monitor temperature stability near the surface of the insole during data acquisition. In another of their article focusing more on the e-health solution part [31], Domingues et al. discussed different aspects of the e-health system more in detail along with a description of their developed FBG-based insole. The multiple set of research conducted by this group has been presented in detail in this book chapter [32]. So, based on the current literature, it is evident that none of the groups so far conducted even a standalone study for foot temperature measurements using FBG sensors regardless of their capability to smartly measure both pressure and temperature simultaneously in the same system, as proven in the current study. Some studies only used FBG-based temperature sensors for calibration or compensation purposes. On the other hand, even though some of the studies showed some basic comparison between optical and electronic sensing units in capturing plantar pressure, a detailed comparison from both pressure and temperature aspects is still missing. While developing a new solution, it is a standard practice to compare the developed system with state-of-the-art, commercial solutions widely used in the domain. Such comparisons are also vastly missing in the current literature which has been performed in the current study.

Therefore, to address all the issues with available options for wearable diabetic foot complication detection, in this paper, we propose to develop an optoelectronic sensor (FBG)-based wearable smart footwear for continuous plantar temperature and pressure measurement systems, which will measure the daily asymmetric temperature and pressure variations due to the gait dynamics for early, reliable, and robust detection of diabetic foot complication. The smart footwear will be used for plantar temperature and pressure acquisition during complete gait cycles when the diabetic patient will be asked to walk while wearing the shoe. The captured temperature and pressure data generated throughout several full gait cycles can be sent to the PC module, logged into a database for future analysis, and can also be used to synchronously update the gait dynamics. In this way, the patient will get an early warning and can contact health care professionals to take preventive measures before the disease reaches chronic conditions.

The paper is divided into 5 sections. Section I reflects the motivation behind the study and some of the recent works done in this area while Section II explains the methodology adopted for the characterization of the FBG sensors for plantar temperature and pressure measurements along with brief descriptions of the other two systems explored in this study for comparison. Section III describes the experimental setups in detail while Section IV provides the study outcomes and explanations. Finally, Section V contains the concluding remarks from this study.

## 2. Materials and Methods

In this section, at first, we discussed briefly on the data acquisition process from a lab-made, low-cost electronic sensor-based smart insole for plantar pressure and temperature measurement developed by our team in a recent study [33]. Next, we give the reader a brief idea about the widely-used F-Scan In-Shoe Plantar Pressure Measurement System for Gait Analysis. Finally, we discuss in detail the developed FBG-based smart insole for gait cycle (foot pressure) and foot temperature data acquisition. This section also introduces the layouts of the insoles, data acquisition system, hardware and software interfaces, and calibration of the sensors and system, respectively for all three systems. In **Figure 1**, a 2D frontal view of both developed smart insoles is shown in comparison to the commercially available state-of-the-art F-Scan system from TekScan[TM] [34].

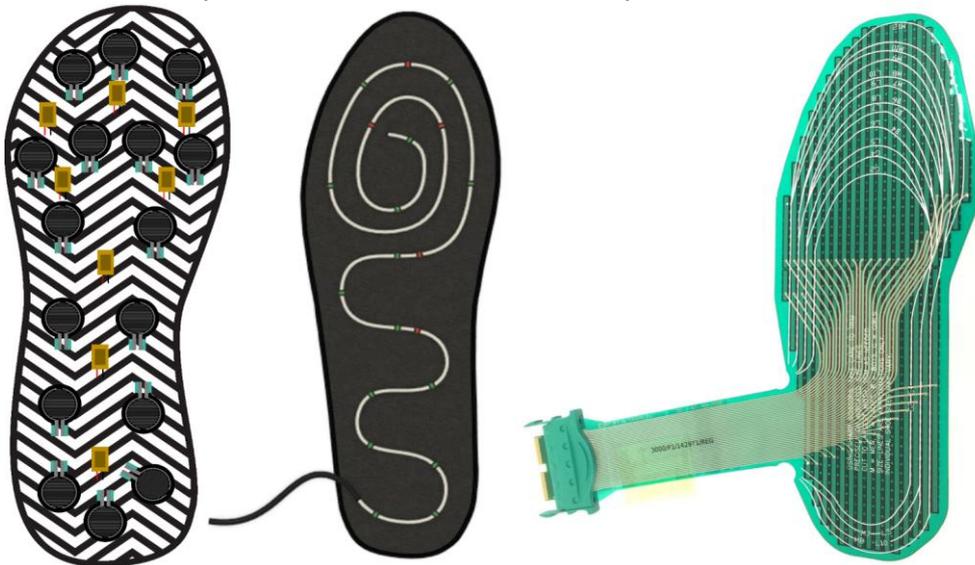

**Figure 1.** Designed smart insoles with electronic sensors (left) [33], Optoelectronic FBG sensors (middle), and commercially available F-Scan smart insole (right) for vGRF data acquisition by TekScan[TM] [34].

### 2.1 Electronic Sensor-based Plantar Pressure and Temperature Acquisition System

This insole utilized electronic sensors such as Force Sensitive Resistors (FSRs) for pressure measurement and Thermistors (NTCs) for temperature acquisition. The development process can be divided primarily into four tasks viz. selection and characterization of pressure and temperature sensors, smart insole design, design of data acquisition system for acquiring and wirelessly transmitting data, and setting up a real-time data logger for saving the acquired data in a local database.

a. Pressure Sensor Characterization

Three different types of pressure sensors (Velostat, Piezo-electric, and Force Sensitive Resistor (FSR)) were investigated and characterized to identify the most suitable sensing unit for the smart insole application. The detailed process of the calibration of all transducers has been reported in Khandakar et al. [33]. As reported in [33], among the three tested pressure sensors, FSR was found suitable for the insole application. The calibration equation can be written as in Equation (1). If we consider FSR is in series with the external resistance $R_{ext}$ and is supplied by the voltage $V_{cc}$, the output voltage from the FSR can be formulated as,

$$V_{out} = V_{cc} * \left(\frac{R_{ext}}{R_{ext} + R_{FSR}}\right) \Rightarrow R_{FSR} = \frac{V_{cc} * R_{ext}}{V_{out}} - R_{ext} = \frac{R_{ext}(V_{cc} - V_{out})}{V_{out}} \qquad (1)$$

According to Interlink Electronics [35], FSR resistance is approximately 10kΩ for an applied force of 100g over the surface of the FSR. But this force is much lower than the mean peak force applied by an adult human in various regions under the feet. Perttunan et al. in their detailed thesis work [36] reported that peak plantar pressure in various regions under the feet of an average adult human varies from 80kPa to around 600kPa. So, the maximum pressure any FSR will face varies in each region under the feet. This data needs to be properly stored as it will be required by any statistical or machine learning (ML) based classification technique to be applied on plantar pressure data to detect foot or other relevant anomalies. To convert this pressure unit to a force or weight unit, we need to consider the "Active Sensing Area" of the FSR units. This relation can be formulated as,

$$\text{Applied Pressure} = \frac{\text{Overall Force Applied}}{\text{Active Sensing Area of FSR}} \qquad (2)$$

The active sensing region for FSR-402 is a circular region with a radius of 6.35 mm and so the area of the sensing region, $A = \pi R^2 \cong 126.677$ mm$^2$ [35]. We know that 1 Pascal (Pa) is equivalent to 1N of force on an area of a one-meter square. So, the force applied to the FSRs embedded in the insoles varies from around $\frac{80000 * 126.677}{1000 * 1000} = 10N$ to $\frac{600000 * 126.677}{1000 * 1000} = 76N$. So, the mass per sensor varies from 1.03kg to 7.75kg. Therefore, this is close to the one-kΩ calibration curve of FSR [35]. From this understanding, we concluded that the external resistance ($R_{ext}$) can be varied from around 600Ω to 1200Ω to produce an optimum response for foot-pressure measurement. In this way, the insole was made invariant to the subjects of various weight classes.

b. Temperature Sensor Characterization

Three types of commonly used temperature sensors have been considered for characterization and calibration viz. Thermostats, Resistance Temperature Detectors (RTDs), and Thermistors. Among them, thermistors were chosen since they could be hooked up easily using the same voltage divider circuit as FSRs and the nominal resistance of thermistors is 10kΩ. On the contrary, RTDs require a Wheatstone Bridge for each sensing unit. Placing eight bridges on the PCB would drastically increase its size. The thermostats are not suitable for this application due to their working mechanism. Moreover, Thermistors have a very high response time, faster than RTDs and Thermocouples which helps in recording fast temperature changes. The foot temperature of a normal person can vary from 15°C to about 37.5°C during summer [37]. For patients with complexities such as DFU, it can be higher [38]. Since the aim is to find the temperature of the feet for both healthy and unhealthy subjects, the suitable working range of the temperature sensor was chosen as 20°C to 50°C. The calibration process is reported in detail in Khandakar et al. [33]. Thermistors are mainly of two types viz. Positive Temperature Coefficient (PTC) and Negative Temperature Coefficient (NTC). The temperature reading from the NTC type was much more reliable and widely used, which can be derived by using Equation (3).

$$\text{Temperature}, T = \frac{B}{\ln\left(\frac{R_{Thermistor}}{R_0 * e^{-\frac{B}{T_0}}}\right)} \qquad (3)$$

Here, $R_0$ is the nominal resistance for a certain temperature while $T_0$ is the (nominal) temperature for that resistance or vice-versa. 'B' is a constant specific to the thermistor material which is equal to 4350K.

c. Smart Insole Design and Sensor Placement

There are 24 sensors (16 FSRs and 8 Thermistors) soldered to a ribbon cable used to connect the insole to the data acquisition module. The sensors had a sequence, which was followed during data transmission, data logging, and even during foot pressure and temperature map generation. Sensor placement locations were selected based on the temperature hot spot [33] and the pressure points important to produce a reliable vertical ground reaction force map or plantar pressure [33]. Supplementary **Figure S1** shows the insole with pressure and temperature sensors along with the compact data acquisition system. The pressure and temperature data are wirelessly sent to the central Bluetooth Low Energy (BLE) device connected to a PC from the peripheral connected to the insole.

d. Data Acquisition System

There are 16 FSRs and 8 Temperature Sensors in each insole. A 16:1 MUX (multiplexer) for the FSR and an 8:1 MUX for the temperature sensors are used in the circuit, which would require only 2 analog-to-digital converters (ADCs). TinyPICO microcontroller [39] was used as the MCU due to its on-board BLE, Wi-Fi module, 14-bit ADC, recharging option, and a very small form factor. Following Equation (4) below, the output from the ADC and the acquired temperature can be related as,

$$\text{Temperature, T} \approx \frac{\ln\left(\frac{40950000 - 10000 * A}{A}\right) - 3.10}{-0.048} \quad (4)$$

where we can get the value of 'A' or the analog reading from the sensor output. This equation has been taken directly from [33], and detailed derivation steps are available there. Note that this formula will only hold when the external resistance used is 10kΩ, for other values it needs to be recalculated.

e. Real-Time Data Logging

Real-Time Data Logger was developed so that the data received by the MCU acting as the Central BLE can be saved permanently in a local database. The Data Logger was developed based on Python and can generate both text and CSV and MS Excel files, details of which are reported in [33]. The data logger dealt with four data streams coming from the data acquisition devices viz. Right Leg Pressure and Temperature, and Left Leg Pressure and Temperature. The data logger was coded to extract incoming data streams and save each subject data into four separate files in a predefined directory.

2.2 F-Scan In-Shoe Plantar Pressure Measurement System for Gait Analysis

The F-Scan is one of the in-shoe plantar pressure measurement systems from TekScan[TM] [34] that provides useful information for diagnosing pathologies, evaluating treatments, educating patients, assessing subjects, etc. from gait cycle measurements. The F-Scan software includes features such as report generation for experiments, patient database, plantar pressure map generation, peak pressure analysis, force vs. time graphs, side-by-side comparison of pre and post recordings, per patient and studies robust calibration, an external triggering option, etc. The recordings can be saved in various formats such as exporting recordings as Audio Video Interleave (AVI) files, saving data as American Standard Code for Information Interchange (ASCII) files or Microsoft Excel tables, etc. F-Scan comes in both wired and wireless options, wired system (3000E VersaTek) was used in this study. Each insole before trimming contains a total of 960 individual sensing units. Two VersaTek Cuffs (VC-1), attached to both legs near the ankle through flexible Velcro[TM] bands receive streaming data from smart insoles, prepare it, and send it to the central hub, VersaTek 2-Port Hub (V2PH-1). V2PH-1 remains connected to the VC-1s through 25 feet CAT5e Cuff Cables and to the PC via a much shorter Universal Serial Bus (USB) cable. F-Scan also comes with software for sensor calibration, data logging, visualization, analysis, etc.

2.3 FBG Sensor-based Developed Plantar Pressure and Temperature Acquisition System

This section focuses on the developed plantar pressure and temperature acquitting system using optoelectronic Fiber Bragg Grated (FBG) sensors. In this section, at first, we briefly review the working principles of Fiber Bragg Grated optical cables. Then we provide brief information about the complete optical interrogator system used in this study. Next, we discuss the FBG characterization process for both pressure and temperature data acquisition. After that, we elaborate on the entire smart insole creation process. We end the section with a brief on the data acquisition and visualization software tools used or developed for this study.

a. Working principle of the Fiber Bragg Grated (FBG) Optical Sensors

Fiber Bragg Gratings (FBGs) are durable lightweight sensors that are made from silica core covered in a plastic jacket. They are made in a section of an optical fiber with a periodic variation in the refractive index of the fiber core in the longitudinal direction (approx. 10nm) of the fiber [40]. When light passes through the FBG, only the light of a particular wavelength is reflected, as shown in **Figure 2(a)**. The central wavelength of the reflected waveform from a specific grating is known as Bragg's wavelength. When the optical fiber is strained or pressurized, the reflected spectrum shifts proportionally, as can be seen in **Figures 2**. Changes in the reflected wavelength also happen due to temperature changes. Different sensors manufactured using gratings with a specific wavelength can be implemented in a series on the same optical line. FBG sensors work in the principle of monitoring the reflected wavelength described by ($\lambda_B$ – Bragg's wavelength) and its variation as a function of the parameter under study (e.g., strain and/or temperature). The Bragg wavelength shift, $\Delta\lambda_B$, is related to temperature and strain as follows [41]:

$$\frac{\Delta\lambda_B}{\lambda_B} = (1 - p_e) \times \Delta\varepsilon + (\alpha + \zeta) \times \Delta T = S_\epsilon \times \Delta\epsilon + S_T \times \Delta \qquad (5)$$

where,

$\Delta\lambda_B$ = Change in the Bragg Wavelength
$\lambda_B$ = The Central or the Characteristic wavelength
$\Delta\varepsilon$ = Change in the Strain along with the whole fiber
$p_e$ = Effective Strain Optic constant
$\zeta$ = Thermo-Optic coefficient
$\Delta T$ = Change in the Temperature
$S_\varepsilon$ = Sensor sensitivities to strain variations
and $S_T$ = Sensor sensitivities to temperature variations

The first term is associated with the strain influence on $\lambda_B$, whereas the second term represents the effect of temperature on $\lambda_B$. From **Figure 2(b)**, it can be noted that ideally there is a linear relationship between the shift of the wavelength due to strain or temperature variation in FBG with the changes in delay, which can be translated into a strain or temperature data if the material constant (slope) is known.

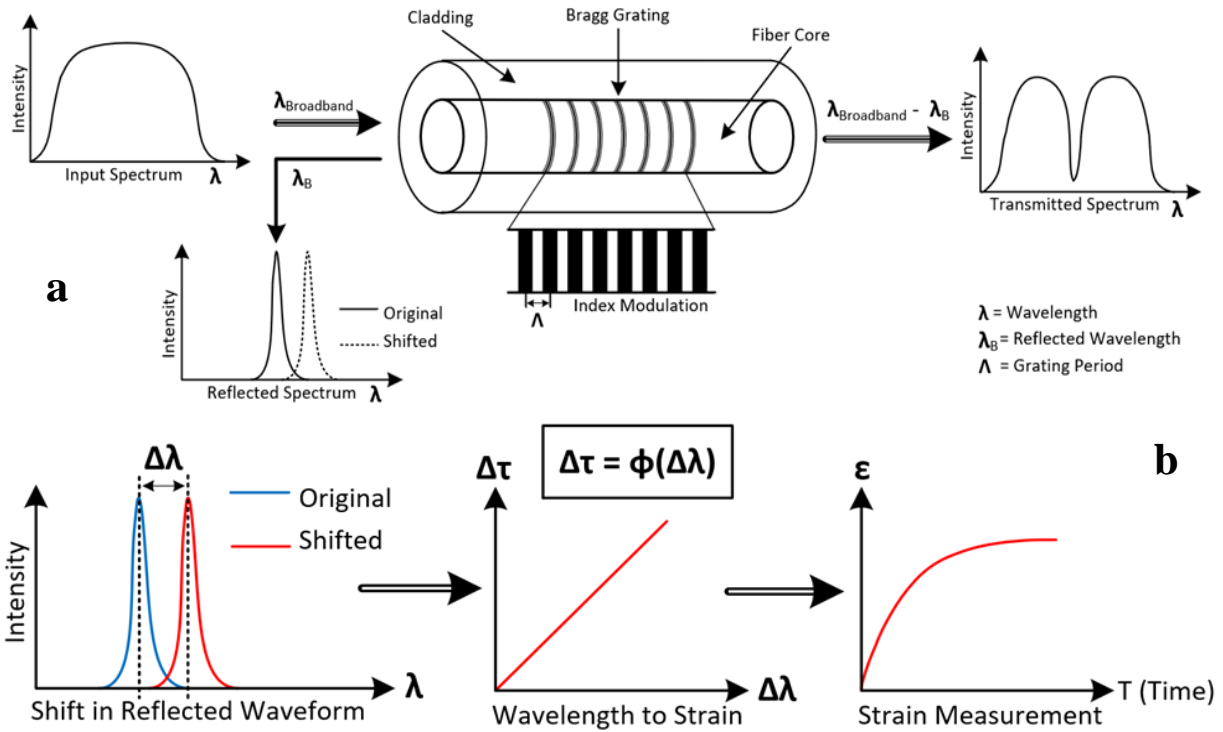

**Figure 2:** Illustration of working principle of FBG sensors (a) and relationship between wavelength shift with the applied strain in FBG sensors (b).

b. FBG Optical Cable and Interrogator Selection

After reviewing several off-the-shelf FBG Sensors, we have selected the FBG sensor array from FiSens[TM] [40]. The advantages of the FiSens sensor array were customizable fiber length, the number of FBG sensing units, spacing of the sensors, etc. making it easier for us to design a model of the insole along with all specifications and ask the company to fabricate the FBG sensor array accordingly. So, it was possible to order a customized fiber optic cable with the pressure and temperature sensors located as per our design. The length of the exposed pure silica fiber over the insole and the polyimide-coated fiber acting as a guard wire were also predefined. The Interrogator used for this work was FiSpec FBG X100, which covers a spectrum of 808-880 nm from FiSens[TM] [40], which can be referred to in supplementary **Figure S2**. Even though it is a tethered system, the interrogator is very compact (65mm×48.5mm×15.3mm) and lightweight (60g) and its high precision in data collection eased the dynamic gait cycle recording process significantly. The module's sampling frequency can be varied between 1 to 100 Hz but precision drops as the sampling frequency are made higher. The interrogator works like an interface between the FBG cable and the PC. It connects the FBG cable with an FC/APC connector and connects the PC with a USB cable. The interrogator module can work with one channel (corresponding to one fiber) only, however, it can handle a maximum of 30 sensors while maintaining its stability, as informed by FiSens[TM]. In our FBG data acquisition system, we placed 15 FBG sensors for pressure and 5 for temperature i.e., a total of 20.

c. FBG Sensor Characterization

Since FBG cables contain the pressure and temperature sensors in series within a single fiber which creates interactions between pressure and temperature sensing units during data acquisition, we have designed the pressure and temperature sensor characterization procedure separately. There are mainly three aspects of FBG characterization viz. pressure characterization, temperature characterization, and pressure or temperature compensation [42].

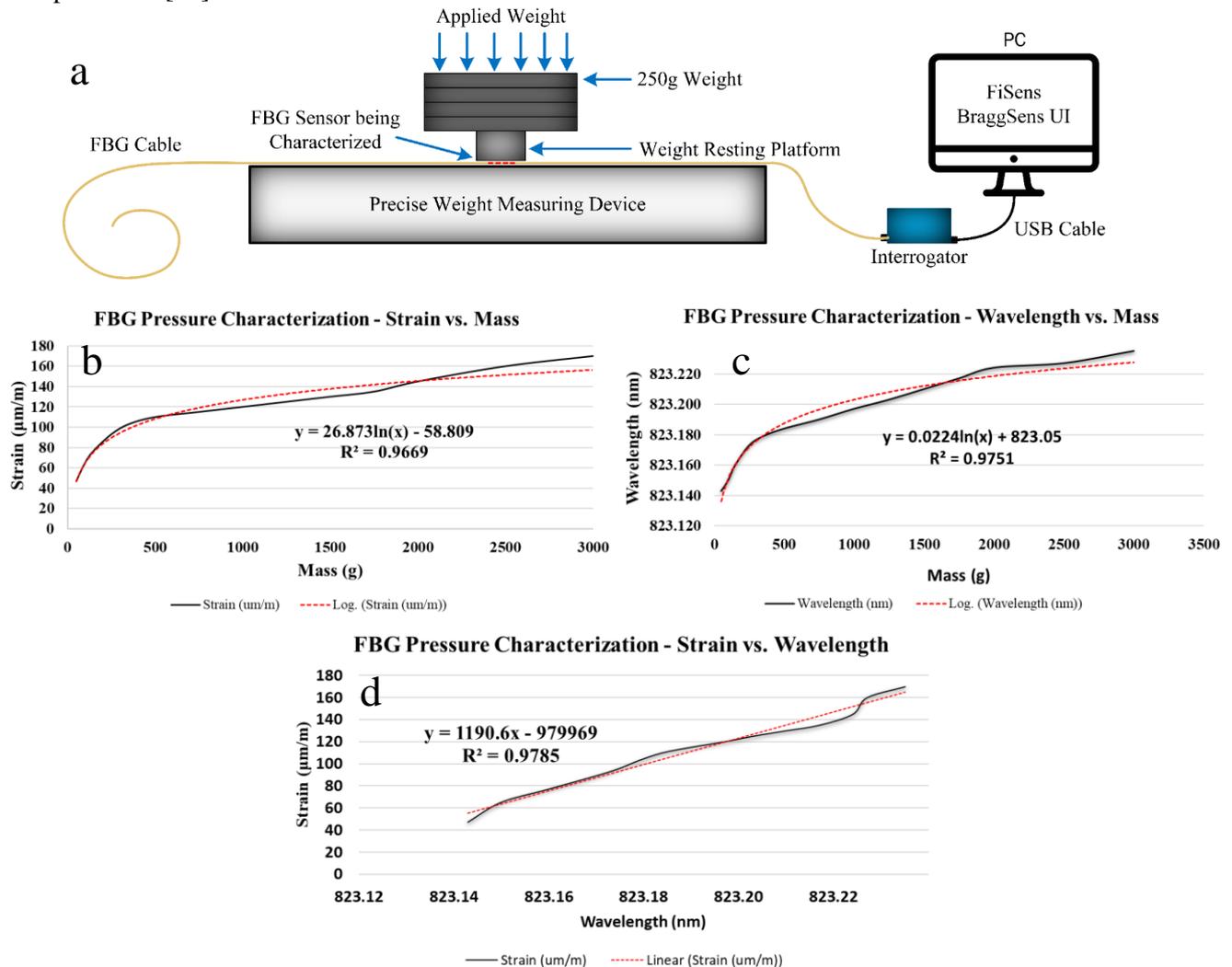

**Figure 3.** (a) Pressure characterization setup for the FBG sensors; FBG sensor's characteristics: (b) Strain vs. Mass; (c) Wavelength vs. Mass and (d) Strain vs. Wavelength shift plot for FBG sensors during pressure characterization.

**FBG Pressure Sensor Characterization:** The FBG pressure sensor characterization setup is shown in **Figure 3(a).** It is similar to the FSR sensor characterization setup proposed in this work [43]. The setup was created on top of a mass measuring device which can measure mass from 2g to 10kg with a precision of 1g over the range. The range was enough for the pressure characterization since vertical ground reaction force (vGRF) at a single point on the insole usually does not cross this range [36, 44]. A weight resting platform, which is a 100g weight itself, was used to place 250g bars on top of each other gradually and take strain and wavelength shift readings for the corresponding weights (mass). A logarithmic trend-line was found to be best fitting to the trend as shown in **Figure 3(b-c)**. Here, the wavelength was measured in nanometer (nm) while the change in strain was reported as micro-meter per meter (μm/m).

Based on the trendline equations, the strain values from the BraggSens software [45] were converted to mass values (equivalent to vGRF) using LabVIEW and saved locally. The strain change measured by FBG corresponds to the deviation in the reflected wavelength from the central wavelength for any sensor. The relation between the wavelength and the weight applied is logarithmic, just like the trend between strain and mass. It validates the characterization process by showing approximately a linear relationship between the wavelength deviation and the change in the strain as reported by the manufacturer and in literature [25, 28, 31] (**Figure 3(d)**).

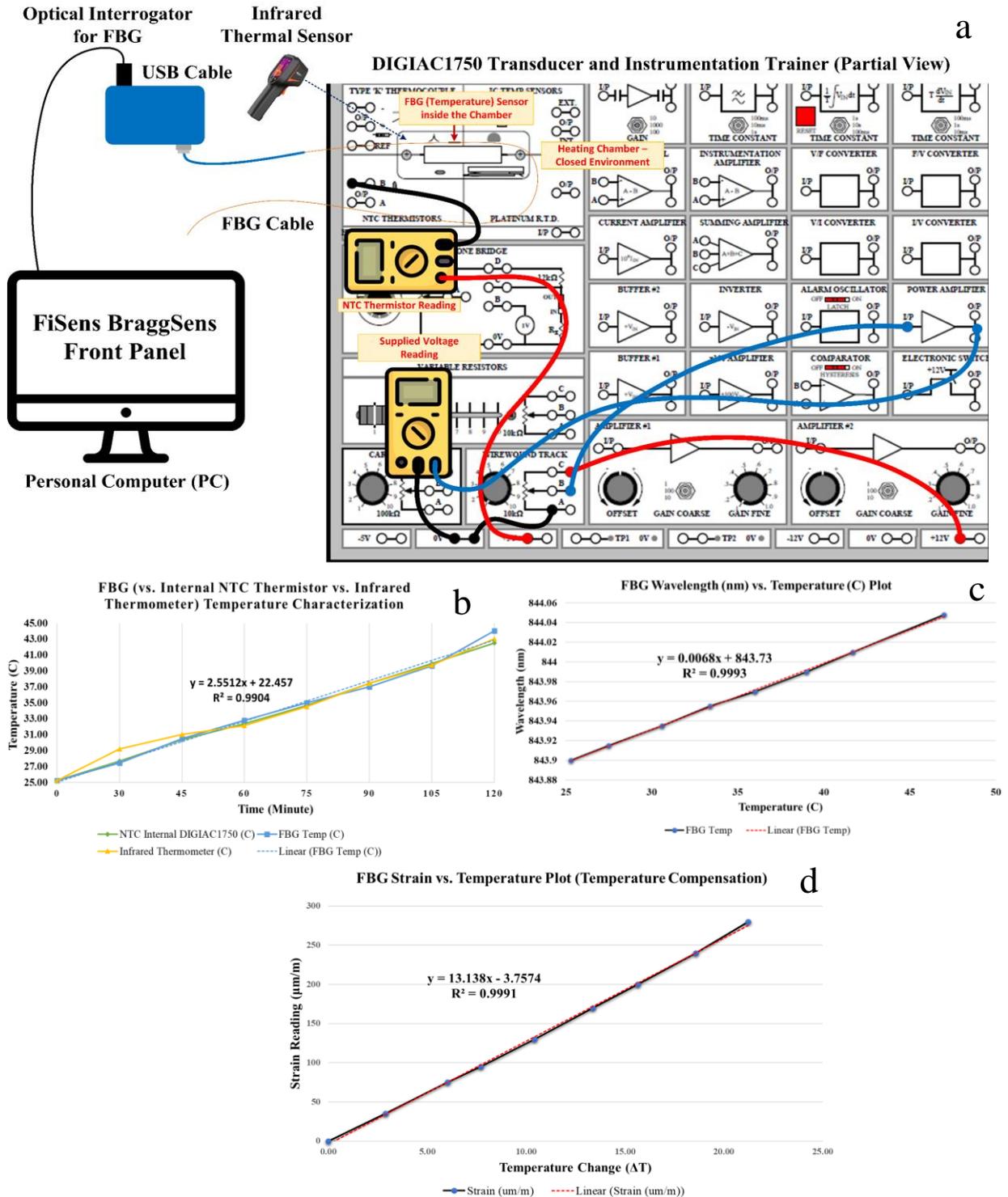

**Figure 4.** (a) Temperature characterization setup for the FBG Sensors; FBG sensor's characteristics: (b) FBG and Reference temperature sensors' response vs. Time, (c) Wavelength shift vs. Temperature for FBG; (d) Temperature compensation (FBG Strain vs. Temperature).

**FBG Temperature Sensor Characterization and Compensation:** The temperature characterization procedure for the FBG sensors was performed using DIGIAC 1750 Transducer and Instrumentation Trainer [46] setup shown in **Figure 4(a)**. The FBG sensor under experiment was put inside an enclosed heating chamber and was heated gradually from a room temperature of 25°C along with two other reference sensors viz. an on-board NTC Thermistor and an external high-performance Infrared Thermometer from FLIR[TM] [47]. The response time there is different, so the heating process was kept constant after some time to let the slower sensor become stable. The Time vs. Temperature plot from BraggSens Front Panel has been shown in supplementary **Figure S3**, which shows how the heating was performed to let all the sensors come into stability due to their response time being different.

Now, there are two aspects of the temperature characterization process viz. Temperature vs. Time for FBG and other reference sensors (**Figure 4(b)**), and Wavelength vs. Temperature response (**Figure 4(c)**). In **Figure 4(b)**, the temperature recorded from the FBG sensor was compared to the temperature recorded by the internal NTC thermistor and infrared thermal sensor with respect to time as the enclosed chamber was heated. The high correlation in their trend is proof of the robustness of both the characterization process and the sensors. The relation between the wavelength shift and temperature change is linear, as shown in **Figure 4(c)**. The plots in **Figure 4** also match the temperature characterization plots in [23]. Based on our literature review, it is the only study that recorded both pressure and temperature simultaneously using FBG sensors for foot condition monitoring, all other studies used a single FBG sensor for temperature compensation.

Since FBG sensors are created in series along the same fiber, measurement of temperature gets affected by pressure and vice-versa [48, 49], especially when the same cable is used to measure both quantities simultaneously. So, while measuring temperature, the effect of pressure is required to be compensated and vice-versa. But from the literature [26, 29, 31], it has been found that the effect of pressure on temperature is negligible, but the opposite is significant. To understand the effect of temperature on pressure for the cables used for our experiment, we recorded the change in strain reading while applying temperature. From **Figure 4(d)**, a linear relationship can be found between the strain reading and temperature change. For each degree of temperature change, a strain change of 13.138 µm/m was observed. From the FBG pressure characterization equation (**Figure 3(b)**) it was observed that such an amount of strain corresponds to approximately 15g of vGRF. It is crucial since, during data acquisition, the foot temperature will rise after a few minutes of walking which might significantly affect the pressure readings from FBG. It is worth mentioning that the BraggSens software provided by FiSens[TM] [40] cannot acquire pressure and temperature readings simultaneously for a single sensor, but by turns. This system can be used for temperature compensation on board (using hardware) or programmatically by setting one or more of the temperature sensors in compensation mode (i.e., as reference). For our experiments, we used all temperature sensors as references during strain measurement and selected the hardware mode for temperature compensation, which was found to be more stable. The FBG interrogator module converts the FBG wavelength shift reading to temperature on board, which minimizes the load on the software (i.e., PC) during signal/data processing.

d. Smart Insole Design and Sensor Placement

The FBG cables and the optical interrogator modules were ordered from FiSens[TM] based on the specifications illustrated by our team simulating the real application, as shown in **Figure 5**. One of the crucial parameters is the minimum bending radius of the fiber cable during usage [50], crossing that limit might snap the fiber. FiSens[TM] set a limit of 10 mm for the minimum bending radius. After a few fiber layout designs attempts over the insole surface, this constraint was met. The minimum bending radius in the design was 14 mm near the central forefoot.

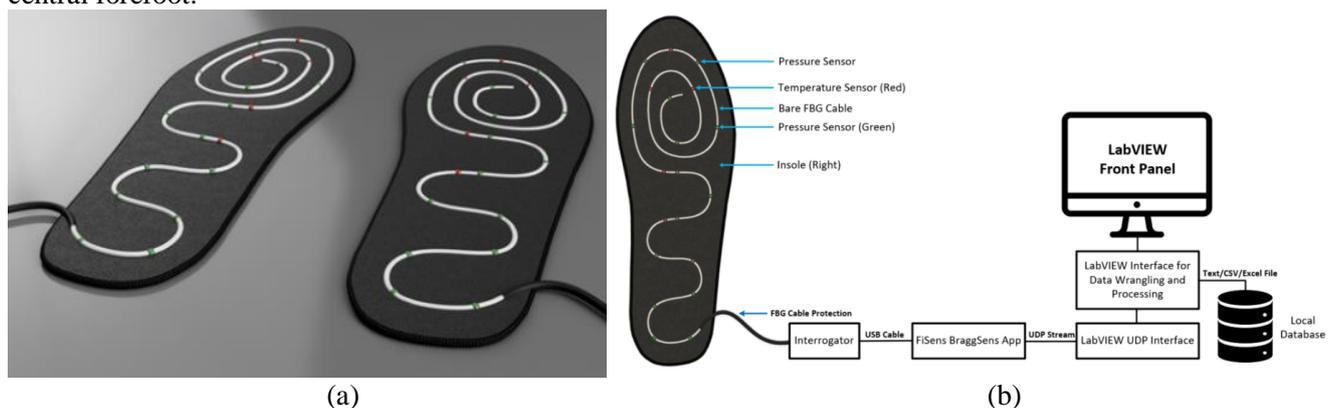

**Figure 5. (a)** FBG insole model in 3D; **(b)** Block diagram for data acquisition system for the FBG insole.

The manufacturer required the exact locations of the FBG sensors on the fiber cable (**Supplementary Table 1**) and the maximum length of the fiber cable with and without the Guard. The manufacturing cost depended on the fiber length, the number of sensors, and some other parameters such as cladding material, integrator type, etc. [45]. It is mentionable that the insole size was chosen as EU-42 Men, the same as the FSR insole, which is also the global average for foot size [51]. So, in the 3D environment, the dimensions (259 mm × 88 mm) of the insole were kept the same as the real insole to estimate the fiber cable length. The orientation of the cable onto the insole surface is shown in **Figure 5**. The main aim of the design was to optimize the locations of the sensors across the entire insole in terms of both pressure and temperature data collection while meeting all constraints from the manufacturer. If noticed carefully, more pressure sensors were placed near the heels and toe regions of the feet due to these locations being more prone to foot complications [52, 53]. If each sensor data is not normalized against the person's Body Mass Index (BMI), it becomes important to maintain the sensor density in the heel and toe regions to avoid getting biased plots in terms of magnitude. Temperature sensors were placed more uniformly across the upper portion of the insole. From the interrogator to the bare fiber, there is a portion of the cable covered by a strong protective material known as the Guard. The protection length was chosen to be 400 mm. Choosing a suitable protection length is important since too short or too long protection might create issues during gait cycle data collection. The length of the entire fiber was around 1030 mm (approx. 1m). As in **Supplementary Table 1**, The sensor number counting starts from the one nearest to the protection and so on. **Figure 5 (b)** properly illustrates the complete setup, explained in details in the next section.

e. Data Acquisition System

Since FBG sensors work based on light instead of electricity, the data acquisition system (**Figure 5(b)**) is largely different from any system built for electronic sensors. One of the most crucial items in this system is the Optical Interrogator. The optical interrogator consists of a complex circuitry containing Tuneable Lasers as light source, Optical Amplifiers for amplifying source signals, Optical Circulators for circulating reflected light bands, photodetectors for capturing reflected lights, etc. [54]. The optical interrogator module provided by the FiSens[TM] contains signal processing circuitry to implement various hardware-based features discussed afterward. So, this module acts as a data acquisition system for multiple FBG sensors connected in series and can work on both static and dynamic applications. The interrogator can measure a large sensing network composed of various types of sensors (such as strain, temperature, displacement, acceleration, tilt, etc.) connected along multiple fibers, by acquiring data simultaneously and at different sampling rates [48]. During data acquisition, the interrogator measures the wavelength associated with the light reflected by the optical sensors and then converts it into the associated physical measurement. As mentioned earlier, an FBG sensor has a central wavelength that gets reflected during normal cases when the light of a broader spectrum is passed through it. One of the common ways to generate such a broadband waveform for an FBG cable is to use a tunable laser [55, 56]. The BraggSens application provided by FiSens[TM] contains a smart Graphical User Interface (GUI) which has mainly two panels viz. Spectrum and Visualization 2D. Along with multiple sub-panels for settings, the Spectrum panel mainly shows the reflected peaks from each sensor while the Visualization 2D panel shows time series plots of the strain response from the sensors. There are two options for peak detection in BraggSens viz. Gaussian (Gauβ) peak detection and onboard peak detection. Peak detection is one of the most crucial tasks for the interrogator since based on the detected peaks, strain and temperature values are measured. Now, through a more convenient software-based Gauβ peak detection method, BraggSens searches for peaks from the received reflected spectrum. On the other hand, if the onboard peak detection option is selected, the hardware tries to estimate the peaks itself shedding some processing loads from the front. But if the hardware method is used, it only sends the peak information to BraggSens instead of the spectrum so the plots are not updated on the display. The UDP Streamflow can be turned ON in BraggSens to capture data in string format (e.g., pressure and temperature sensor data) and send it to our developed LabVIEW UDP Interface, which facilitates the user to log, display, and perform advanced analysis on the data. LabVIEW file can receive the target sensor data, collect and filter it and send it for a display to the front panel and save the data locally. A Butterworth lowpass filter of order 1 and a cutoff frequency of 20 Hz (half of the original 40 Hz sampling frequency by BraggSens) for the gait cycle plots from individual sensors and 10 Hz for the combined output. BraggSens provides the strain values which need to be approximately converted to mass values based on the FBG pressure characterization equation shown in **Figure 3(b)** (Equation (6)).

$$y = 26.873 \ln(x) - 58.809 \Rightarrow x = e^{\left(\frac{y+58.809}{26.873}\right)} \quad (6)$$

Here, 'x' is the mass, and 'y' is strain. Equation 6 was used to convert strain values to mass in real-time in the LabVIEW environment using LabVIEW MATLAB extension to generate vGRF maps afterward using MATLAB Graphical User Interface (GUI). Mentionable that each FBG sensor had its characterization equation for both pressure measurements. The constants varied slightly across sensors and after a long time of usage, it might be required to recalibrate the sensors. For temperature measurements, BraggSens directly provides an estimated value based on the characterization equations set by the manufacturer.

## 3. Experiments

In this section, we describe the experimental setup prepared for plantar data acquisition using the F-Scan In-Shoe Plantar Pressure Measurement System and our developed FBG-based plantar pressure and temperature data collection system.

### 3.1 Experimental Setup for the F-Scan System

F-Scan smart insoles have been trimmed down from the default XL size to approximately match the foot size for the subject under study (**Supplementary Figure S4 (a)**). Even though the number of sensing point decrease from the original 960 after trimming, the sensor density remains the same for smaller insoles (**Supplementary Figure S4 (b)**). This makes the insoles suitable for humans of various foot sizes as they can be permanently trimmed down to match a range of smaller foot sizes suitable for average human adults. The subject wore socks during data acquisition so that the system does not damage, as recommended by TekScan[TM] [34]. Two VC-1 VersaTek Cuff were attached to both legs near the ankle of the subject through flexible Velcro[TM] bands (**Supplementary Figure S4 (c)**), receive streaming data from smart insoles, prepare it, and send it to the central hub, VersaTek 2-Port Hub (V2PH-1). All the hanging cables are attached to the waist with the help of a waist belt to boost comfortability during data acquisition through this wired setup (**Supplementary Figure S4 (d, e)**). During data acquisition, after the hardware setup is prepared, patient demographic data is put into the F-Scan software. The first step of data acquisition is calibration which was done for each person and the calibration data is saved locally. As recommended by the manufacturer, recalibration might be needed to perform for an individual if the data is gathered after a month at least. After calibration, the gait cycle can be recorded by letting the subject walk within the range of the cables. Plantar pressure maps for both feet can be generated in real-time, as shown in **Figure 6**. Other analyses such as Multi-Peak Stance can be performed offline. It is necessary to recalibrate each subject before data acquisition, even better during each study for better vGRF approximation during the gait cycle [57].

### 3.2 Experimental Setup for the FBG System

Our FBG system can acquire both pressure and temperature data, unlike F-Scan. Even though the same setup can acquire both kinds of data, the data acquiring steps are slightly different due to the nature of the physical quantities. As in **Figure 5** and **Supplementary Table 1**, the pressure sensors were marked in green in the design while the temperature sensors were marked in red, along with their respective locations in the cable. The FBG cable used for this purpose has 20 sensors and each of them has a central wavelength which is tabulated in **Supplementary Table 1** along with the respective sensor number (Sensor 0 is the first sensor from the protection side and sensor 19 is the last). There are 21 peaks in the spectrum panel as can be seen in **Figure S5(f).** The first peak is far from other peaks, which represents a reference signal to verify the presence of the interrogator. All the sensors' central wavelengths are spread within the 808-880 nm range in an equidistant fashion (**Supplementary Table 1**). The limitation of a maximum of 30 sensors in a cable mentioned earlier is due to the interrogator having many sensors in one cable will not be able to allocate enough space for individual sensor spectrum shifting wavelengths during any perturbation due to the limited spectrum. Both pressure and temperature are measured based on Equation (5) where the amount of wavelength shift is mapped into the change in temperature. One sensor cannot produce both data at the same time due to the effect of the temperature on the pressure reading and vice-versa. Within the BraggSens Visualization 2D Panel (**Figure S5(c)**), it is possible to set any sensor in Pressure or Temperature mode. For example, our insole had five temperature sensors namely, sensor numbers 6, 10, 13, 16, and 17 (**Supplementary Table 1**). During gait cycle recording, the subject was instructed to walk wearing the insole put inside a shoe (**Figure S5(a, b, d, e)**) while the temperature sensors were set in the temperature mode for compensation while others were set in the strain mode for capturing pressure readings. During temperature measurement, all sensors were turned off except the temperature sensors to minimize cross-talk. Since pressure effect on temperature is minimum as explained earlier, there is no option for pressure compensation in BraggSens. To make the temperature compensation robust, pressure sensors were covered by Kapton tapes (**Figure S5(a, b)**) which are normally used as insulators in electrical appliances [58]. Tapes also protected the sewed fiber from snapping during data acquisition. Moreover, the subject was instructed to sit during temperature measurement to further reduce cross-talk among sensors from dynamic movements. **Figure S5(d)** shows the complete wired acquisition system setup (hardware and software). The interrogator was tied to the leg using a flexible Velcro[TM] fastener and the subject wore socks during data acquisition to protect the fibers (**Figure S5(d)**). The experimental setup for the electronic insole is similar to the FBG setup, except that insole is wireless.

## 4. Results and Discussion

In this section, the outputs from the F-Scan and FBG insole systems will be displayed and analyzed for comparison from various technical and economical viewpoints. Outcomes from the electronic insole can be found in [33]. **Figure 6** below shows the output from the F-Scan device. It shows the heat maps for both right and left feet during various stages of the gain cycle. It also shows the Force vs. Time i.e., Gait Cycle plots generated after processing the recorded data.

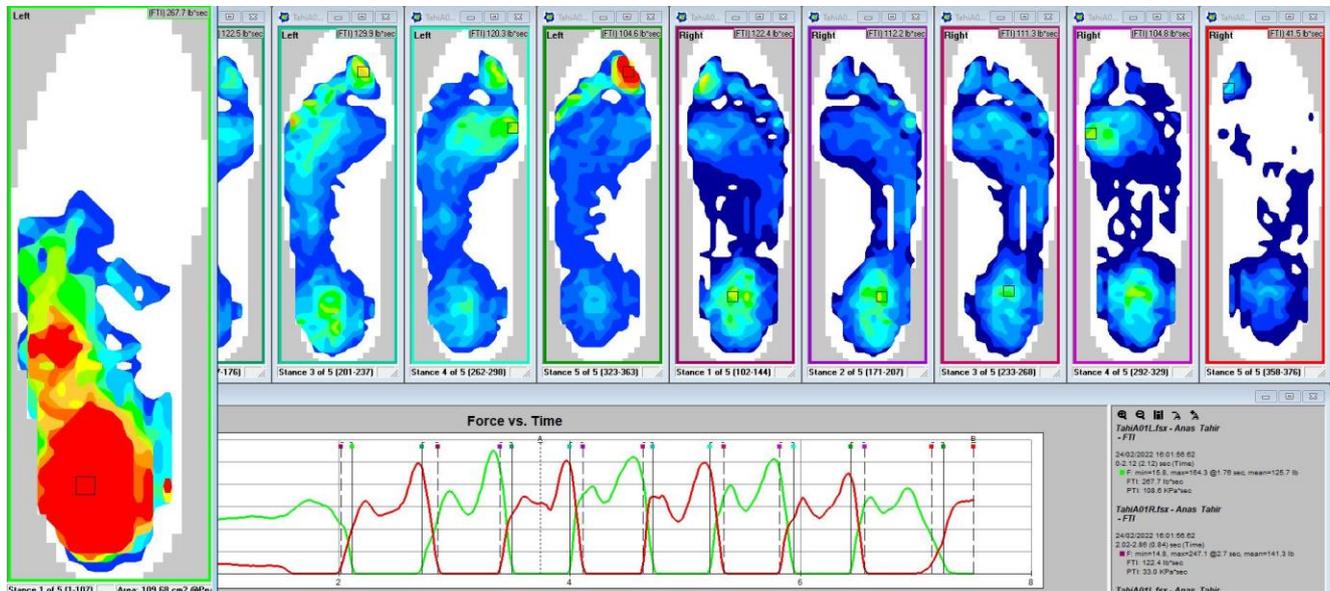

**Figure 6.** Real-Time Heatmaps of both feet for various stances during walking along with Force vs. Time i.e., Gait Cycle plots generated afterward using F-Scan. Here, the red plot represents data from the left leg and the green plot represents data for the right leg.

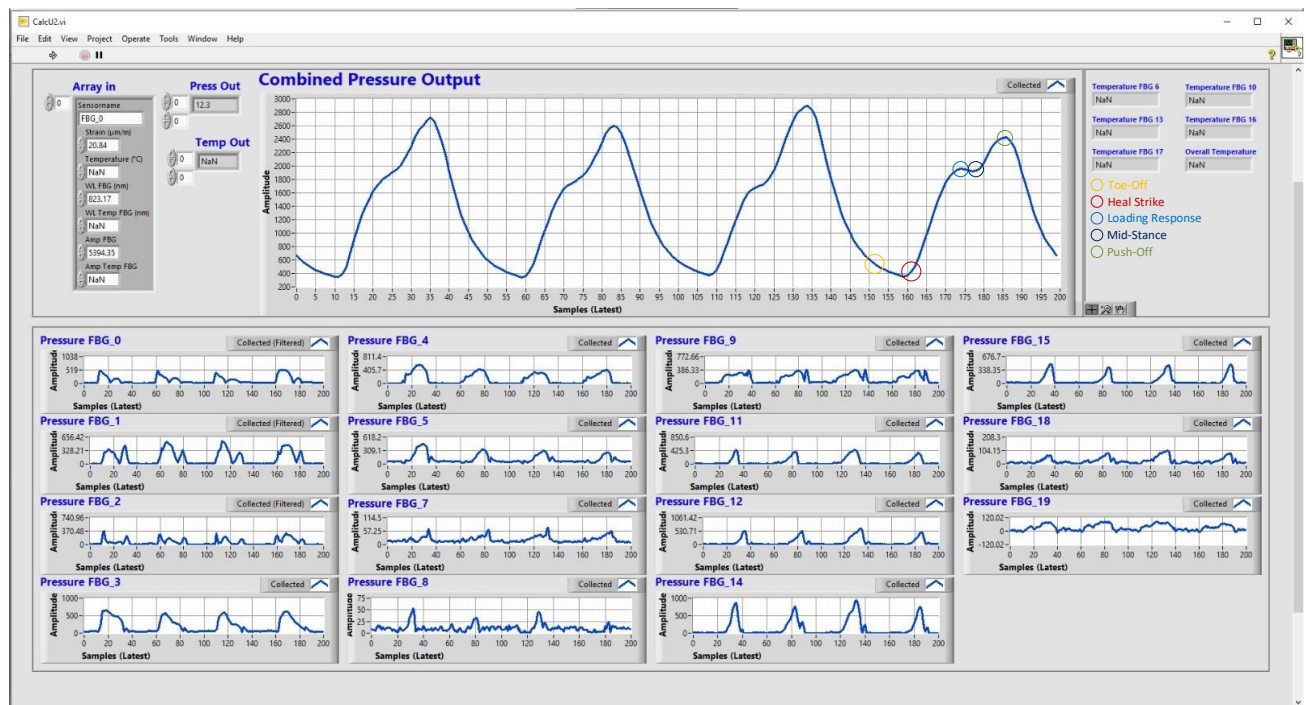

**(a)**

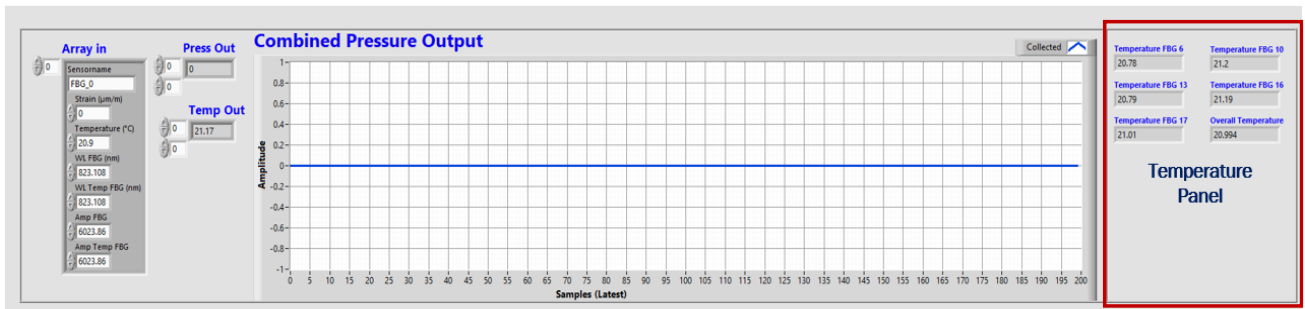

**(b)**

**Figure 7.** (a) vGRF Plots from the FBG Sensors in the Pressure Panel (Temperature Sensors in Compensation Mode); (b) Temperature Values from the FBG Sensors in the Temperature Panel (Red box).

As it can be seen, the pressure maps and gait cycle plots from the F-Scan system are highly accurate, precise, and reliable due to having a robust system that contains so many sensing units on the insole. On the other hand, in **Figure 7(a)**, we display the gait cycle plot on the LabVIEW front panel from our FBG-based smart insole. The combined plot is the summation of the values from all sensors. Here, one of the Gait Cycles was annotated to label various phases of the cycle. The corresponding responses for each FBG sensor have also been plotted. In **Figure 7(b)**, the activated temperature panel has been shown. It is to be remembered that during pressure data collection, the temperature sensors are in compensation mode so they are shown as 'NaN'. On the contrary, during temperature data acquisition, the pressure sensors are turned off in the BraggSens interface to avoid crosstalk among sensors.

In **Figure 8**, the foot pressure map generated from the collected FBG sensor data has been shown. **Figure 8(a)** shows the sensor matrix in MATLAB, **Figure 8(b)** shows the 3D foot pressure map generated while **Figure 8(c)** shows 2D foot pressure maps at various stages of the gait cycle. The pressure maps, even though not as precise as the ones from F-Scan, can display heatmaps with enough precision.

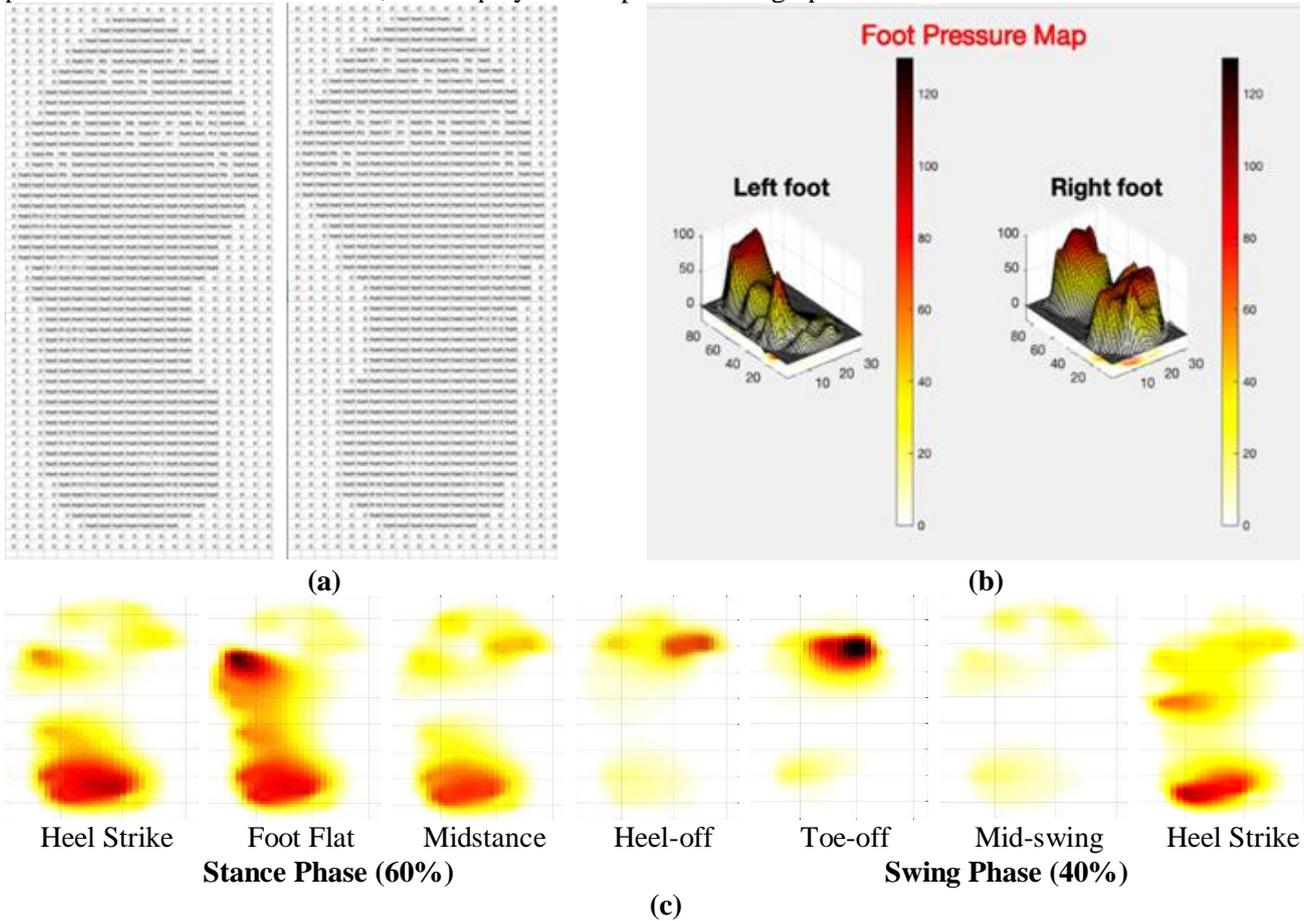

**Figure 8**: (a) Right and left foot template or matrix; (b) 3D Foot Pressure Map; (c) Foot Pressure Maps for various stances during a single Gait Cycle.

**Figure 9** shows the temperature maps from the insole with 5 FBG sensors for temperature measurement during a few stages of the data recording, which shows that foot temperature rose as time passed. The real-time temperature map also displays the foot region with the most temperature in general and the regions which get heated most with time [59, 60, 61]. The subject under test was non-diabetic, the symmetry between the temperature heatmaps of two feet asserts that fact.

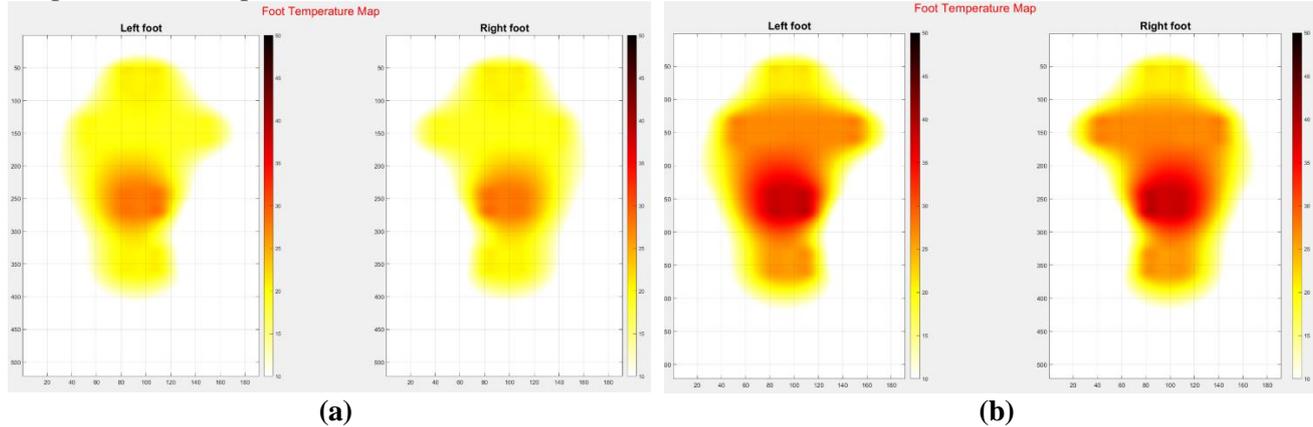

**Figure 9:** Heat map showing Foot Temperature (a) Initial; (b) After around 1 Hour.

The temperature and pressure maps from the electronic sensor are similar, different only due to the sensor locations. Nevertheless, the three insoles discussed in this study differ in many aspects. A comparison chart mentioning all their major differences is provided in **Table 1**.

**Table 1:** Comparison Chart for the three insoles

|  | **FBG Insole** | **Electronic Insole** | **F-Scan System** |
|---|---|---|---|
| **Insole Type** | Research | Research | Commercial |
| **Measured Quantity** | Pressure, Temperature | Pressure, Temperature | Pressure |
| **Number of Sensing Units** | 16 Pressure, 8 Temperature | 15 Pressure, 5 Temperature | 960 Pressure |
| **Calibration** | Automated | None | Manual |
| **Portability** | Wired | Wireless | Wired |
| **Communication Protocol** | Serial | Bluetooth Low Energy (BLE) | Serial |
| **Precision** | Good | Good | Very High |
| **System Robustness** | High | Medium | Very High |
| **Comfortability** | Very High | Medium | High |
| **Fragility** | High | Low | Low |
| **Cost** | High (~2500 USD) | Medium (~350 USD) | Very High (20k USD) |

## 5. Conclusion

In conclusion, in this paper, we have proposed a Fiber Bragg Grated (FBG) optoelectronic insole which is a robust solution for real-time foot pressure and temperature measurement, gait cycle waveform plotting, and heat map generation. While the system can be vulnerable due to the fragile nature of fiber cables, it can be made sturdier with proper manufacturing techniques. The high cost of the developed system is due to the FBG interrogator units from another manufacturer but can be reduced if mass-produced for making insoles. It is still much less costly than the F-Scan system which can only measure pressure. The system can be upgraded to wireless very easily by using the custom UART ports to serially send the data to a microcontroller device and transmit the data wirelessly using communication protocols such as Bluetooth. The system is also comfortable, portable, and lightweight due to the very simple physical nature of the system. While a single fiber cable can work as both pressure and temperature sensors by reducing crosstalk through hardware/software compensation techniques and turning ON/OFF sensors, a more efficient approach will be to use a separate cable for both purposes. The system can reach a high amount of precision even with a few numbers of sensors, as discussed earlier. But one of the crucial challenges will be to increase the number of sensors within the bandwidth constraint. While the number of sensors cannot cross a certain limit within a fixed bandwidth due to a minimum bandwidth allocation for each sensor, the bandwidth can be increased while maintaining efficiency. As a whole, we expect this system to become a useful alternative in the field of diabetic foot monitoring and other similar biomedical applications.

## Acknowledgment

This work was made possible by Qatar National Research Fund (QNRF) NPRP12S-0227-190164 and International Research Collaboration Co-Fund (IRCC) grant: IRCC-2021-001 and Universiti Kebangsaan Malaysia under Grant GUP-2021-019 and DPK-2021-001. The statements made herein are solely the responsibility of the authors. Open access publication is supported by Qatar National Library (QNL).